\documentclass{PoS}
\pdfoutput=1

\title{Study Of Boosted W-Jets And Higgs-Jets With the SiFCC Detector}

\ShortTitle{Study Of Boosted W-Jets And Higgs-Jets With SiFCC}

\author{\speaker{Shin-Shan~Yu},$^{a}$ Sergei~Chekanov,$^{b}$ Lindsey~Gray,$^{c}$ Ashutosh~Kotwal,$^{c,d}$ Sourav~Sen,$^{d}$ and Nhan~Viet~Tran$^{c}$\\ \\
        \llap{$^a$}Department of Physics, National Central University, Chung-Li, Taiwan\\
        \llap{$^b$}High Energy Physics Division, Argonne National Laboratory, Argonne IL, USA\\
        \llap{$^c$}Fermi National Accelerator Laboratory, Batavia, USA\\
        \llap{$^d$}Department of Physics, Duke University, Durham NC, USA\\\\
        E-mail: \email{syu@cern.ch},\email{chekanov@anl.gov},\email{lagray@fnal.gov},
        \email{kotwal@phy.duke.edu},\email{ss567@phy.duke.edu},\email{ntran@fnal.gov}}

\abstract{
We study the detector performance in the reconstruction of 
hadronically-decaying W bosons and Higgs bosons at very high energy proton 
colliders using a full \textsc{Geant4} simulation of the SiFCC detector. The 
W and Higgs bosons carry transverse momentum in the multi-TeV range, which 
results in collimated decay products that are reconstructed as a single jet. 
We present a measurement of the energy response and resolution of 
boosted W-jets and Higgs-jets and show the separation of two sub-jets 
within the boosted boson jet.}

\FullConference{38th International Conference on High Energy Physics\\
		3-10 August 2016\\
		Chicago, USA}

\newcommand{\pt}{\ensuremath{p_{\mathrm{T}}}}
\newcommand{\ptj}{\ensuremath{p_{\mathrm{T}}(\mathrm{jet})}}
\newcommand{\PYTHIA} {{\textsc{pythia}}}
\newcommand{\GEANTfour} {{\textsc{Geant4}}}
\newcommand{\MADGRAPHAMC} {\textsc{MadGraph5\_aMC@NLO}}

\begin{document}

\section{Introduction}
A very high energy proton-proton collider in the future, such as FCC-hh 
at $\sqrt{s}=100$~TeV~\cite{FCC-hh} or 
SPPC at $\sqrt{s}=70$~TeV~\cite{SPPC-one,SPPC-three}, 
will set many challenges for the detector design. High-energy $pp$ colliders 
can produce particles with large mass, giving decay products with large 
transverse momentum. For searches or measurements that involve hadronic 
final states, the response and resolution of jets are major sources of 
systematic uncertainties in the analysis. 
In order to determine precisely the energy of a 
boosted jet, a calorimeter for future $pp$ colliders must satisfy the 
following requirements: (i)
 good containment up to $\ptj\approx 30$~TeV, which implies a total 
 interaction length of $\approx 12 \lambda_\mathrm{I}$ for the full 
 calorimeter (electromagnetic and hadronic)~\cite{bitch};
 (ii) small constant term in the energy resolution of hadron calorimeter: 
$c < 3\%$, since the constant term dominates jet energy resolution for 
$\ptj>5$~TeV;
 (iii) sufficient transverse segmentation for resolving boosted particles;
 and (iv) longitudinal segmentation for detailed information on the shower 
profile.
A good starting point and a promising approach for high energy $pp$ colliders, 
aiming to satisfy the requirements listed above, 
can be a detector based on the Silicon Detector (SiD) concept~\cite{sid} 
developed for the International Linear Collider~\cite{ILCone,ILCtwo}. The 
name of this detector is ``SiFCC'', where ``Si'' indicates that it 
shares several design features of the all-silicon SiD detector. 
This document studies the performance of the SiFCC detector using 
a full \GEANTfour\ simulation and the physics events containing 
boosted W-jets and Higgs-jets. The studies shown in this document 
focus on the performance of the electromagnetic and hadron calorimeters
 of the SiFCC detector.

\section{Description of the SiFCC Detector} 
As with most of the multi-purpose detectors, the SiFCC detector has a 
cylindrical shape. The diameter and the length of the SiFCC detector are 18~m 
and 20.1~m, respectively. The size of 
the SiFCC detector is smaller than the ATLAS detector but  
larger than the CMS detector. 
The version of the SiFCC detector for the study in this document 
is version 7, currently with the focus on the barrel region.
%Figure~\ref{fig:sifcc} illustrates the geometrical sizes of the SiFCC 
%detector.
%
The central feature of the SiFCC detector is a superconducting solenoid, 
of 4.8~m internal diameter, providing an axial magnetic field of 5~Tesla 
along the beam direction. Within the superconducting solenoid volume are 
a silicon pixel and strip tracker, a silicon-tungsten electromagnetic 
calorimeter (ECAL), and a scintillator-steel hadron calorimeter (HCAL). 
The silicon pixel detector has a readout pitch size of 20~$\mu$m, with
5 layers in the barrel and 4 disks in each endcap. 
The silicon strip detector has a readout pitch size of 50~$\mu$m, with 5 
layers in the barrel and 4 disks in each endcap.
 The ECAL, which surrounds the tracker volume, is finely-segmented, and 
 has a cell size of 2~cm$\times$2~cm, 32 longitudinal layers, and a 
total radiation length of $\approx 35 X_0$. The HCAL cell size is 
5~cm$\times$5~cm, with 64 longitudinal layers, and a total interaction 
length of $\approx 11.25 \lambda_{\mathrm I}$ to provide good containment 
for single hadrons with energies above 1 TeV and a \pt= 40~TeV 
jet~\cite{bitch}. 

\section{Simulated Samples \label{sec:sample}}
In order to study the performance of the SiFCC detector, we use 
three different sets of simulated samples: (i) single pion, (ii) 
$\mathrm{Z}^{\prime}\rightarrow \mathrm{WW} \rightarrow \mathrm{q}\bar{\mathrm{q}}^\prime\mathrm{q^\prime}\bar{\mathrm{q}}$, and (iii) 
$\mathrm{H}_2\rightarrow \mathrm{hh} \rightarrow \mathrm{b}\bar{\mathrm{b}} \mathrm{b}\bar{\mathrm{b}}$ in the 2HDM model. 
The single pion samples are generated using the ProMC 
package~\cite{PROMC}. Pions in this sample are 
produced at the origin $(x,y,z)=(0,0,0)$ with a uniform distribution 
in $\phi=0-2\pi$ and $\left|\eta\right| < 3.5$ and the energy of the pion is 
$2^n$ GeV for $n=1-15$. 
Given that our focus is on the detector performance, in order to eliminate 
the need of reconstructing particles from the underlying event present in $pp$ 
collisions, 
both the $\mathrm{Z}^{\prime}$ and the $\mathrm{H}_2$ are produced in 
$\mu^+\mu^-$ collisions. 
The $\mathrm{Z}^{\prime}$ samples are generated with 
\PYTHIA~6~\cite{PYTHIA} in $\mu^+\mu^-$ collisions for 
$\sqrt{s}=M_\mathrm{Z^\prime}=5,10,20,$ and 40~TeV, respectively. 
The 2HDM $\mathrm{H}_2$ sample is generated at LO with 
\MADGRAPHAMC~5.2.3.3~\cite{MADGRAPH} in $\mu^+\mu^-$ collisions at 
$\sqrt{s}=5$~TeV for $M_\mathrm{H_2}=5$~TeV; the parton showering and 
hadronization are performed with \PYTHIA~6. 
All samples are processed through a \GEANTfour-based~\cite{GEANT4} simulation 
of the SiFCC detector version 7. 
The simulated samples are provided by the HepSim repository~\cite{hepsim}. 
and the computations were performed using the Open-Science grid~\cite{osg}. 

%The main strength of 
%this software lies in the fact that it can easily be configured using XML 
%option files, and it has a platform-independent reconstruction which can be 
%easily deployed on computers with different operating systems. 

\section{Energy Response and Resolution for Single Pions}
\begin{figure}
\begin{center}
\begin{tabular}{cc}
\includegraphics[width=.5\textwidth]{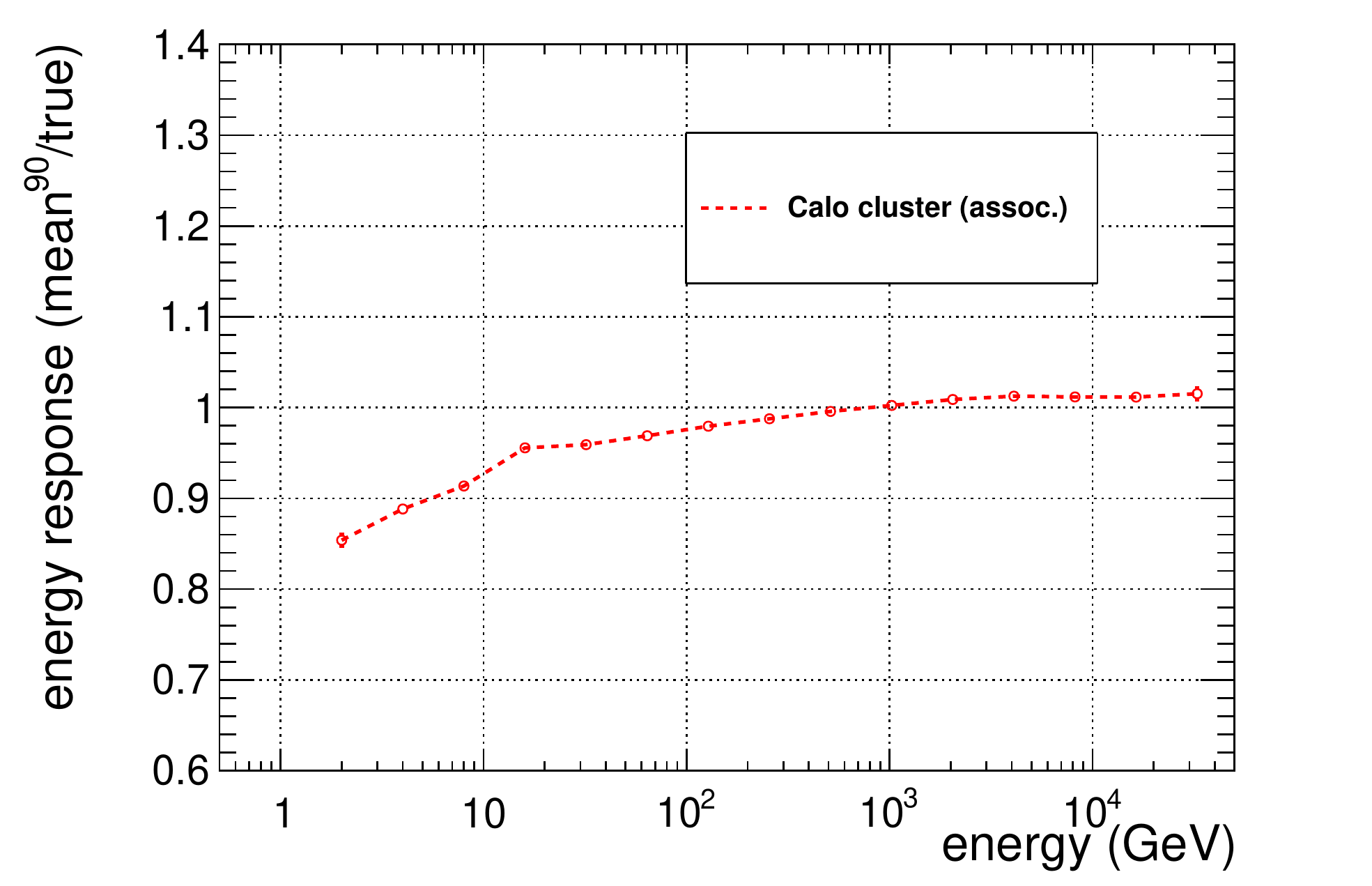} & 
\includegraphics[width=.5\textwidth]{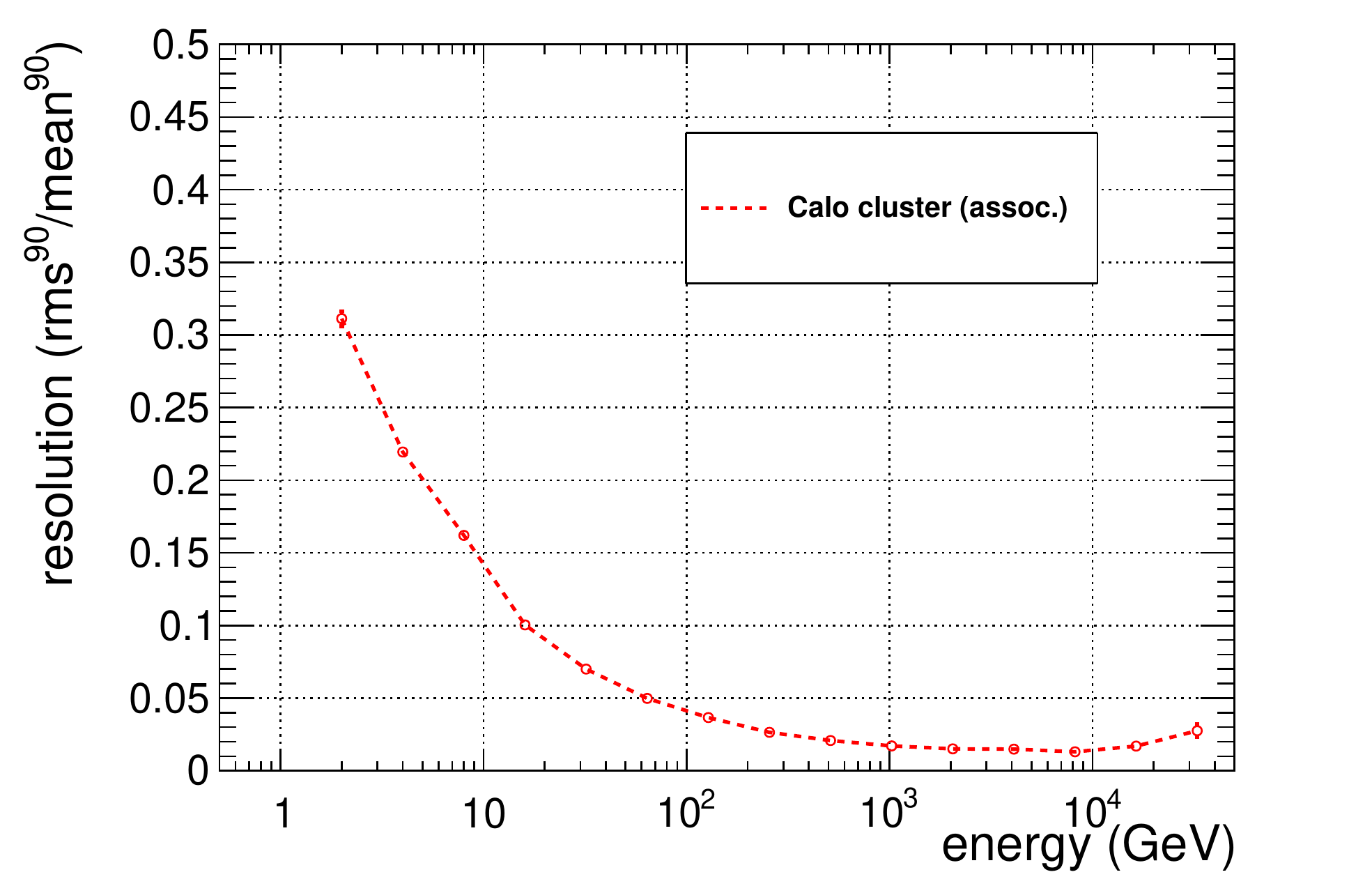}\\
\end{tabular}
\end{center}
\caption{The calorimeter energy response (left) and resolution (right) of a 
single pion as a function of true pion energy. %The solid lines indicate fits to 
%the data points. %The response and resolution using only information from 
%the tracker are also shown for comparison. The calorimeter energy resolution 
%is better than the tracker-only resolution for energy above 2~TeV. 
}
\label{fig:piresponse}
\end{figure}

We study the energy response and resolution of SiFCC calorimeters using the 
single pion samples as described in Section~\ref{sec:sample}. The $\left|\eta\right|$ of the pion must be less than 1.1. The energy 
response is evaluated by taking the 90\% truncated mean of the energy 
clustered with calorimeter hits and dividing it with the true pion energy, 
while the resolution is the ratio of the 90\% truncated RMS to 
the 90\% truncated mean. 
The clustering is performed with the anti-kt algorithm with a distance 
parameter of 0.4~\cite{Cacciari:2008gp}, as implemented in the \textsc{FASTJET} 
package~\cite{fastjet}. 
The calorimeter energy response and resolution in 
Fig.~\ref{fig:piresponse} are fitted with the functions below:
\begin{equation}
 y_\mathrm{response} = A + B\cdot\log\left(E_\mathrm{true}\right), ~~y_\mathrm{resolution} = \frac{0.43}{\sqrt{E_\mathrm{true}}} \oplus 0.009.
\label{eq:response}
\end{equation}
where $A=0.7,0.86,1$, and $B=0.077,0.0202,0$ for $E_\mathrm{true}$ in the 
range of 2--16, 16--2000, and above 2000 GeV.

\section{Weighted Energy Response \label{sec:weight}}
\begin{figure}
\begin{center}
\begin{tabular}{cc}
\includegraphics[width=.4\textwidth]{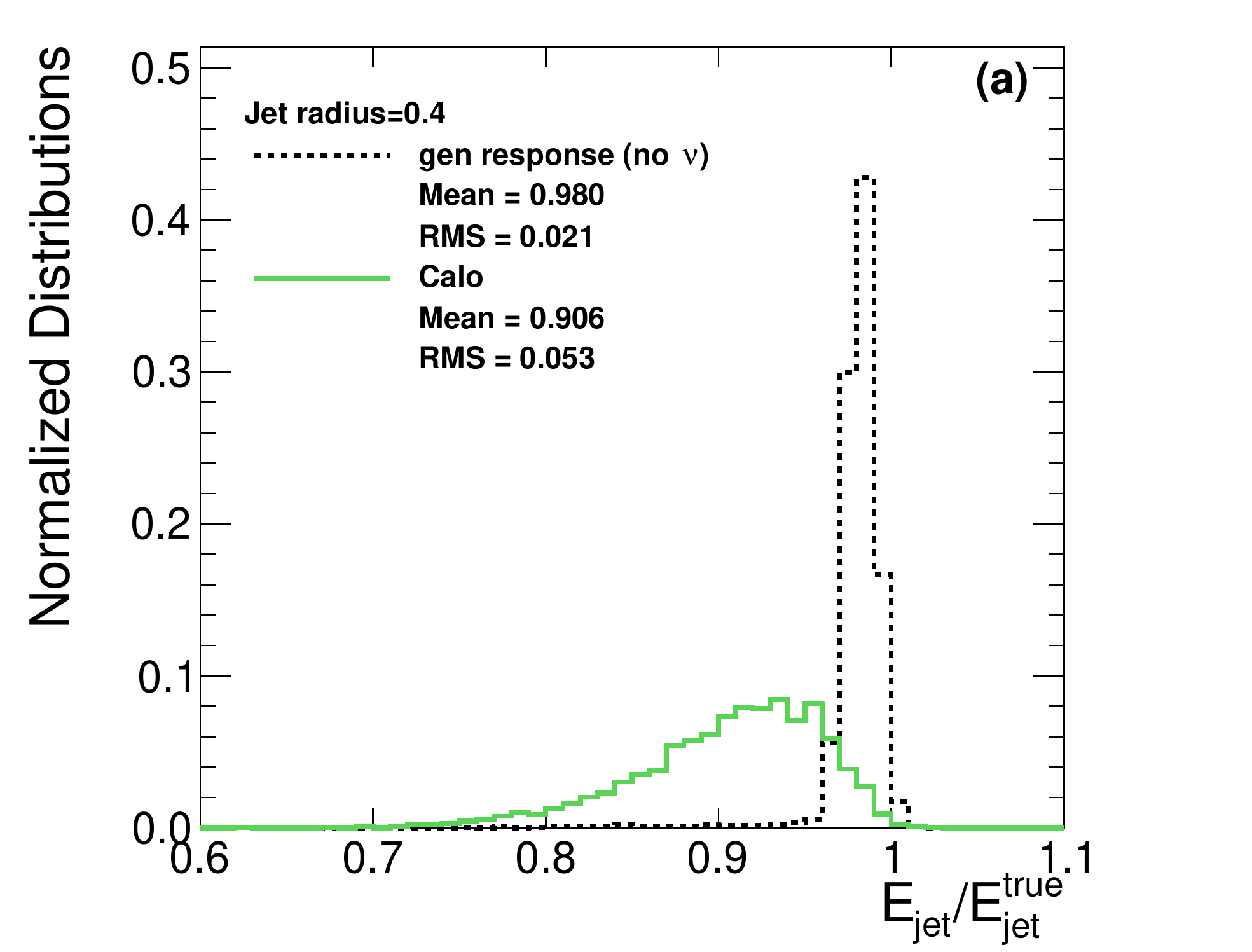} &
\includegraphics[width=.4\textwidth]{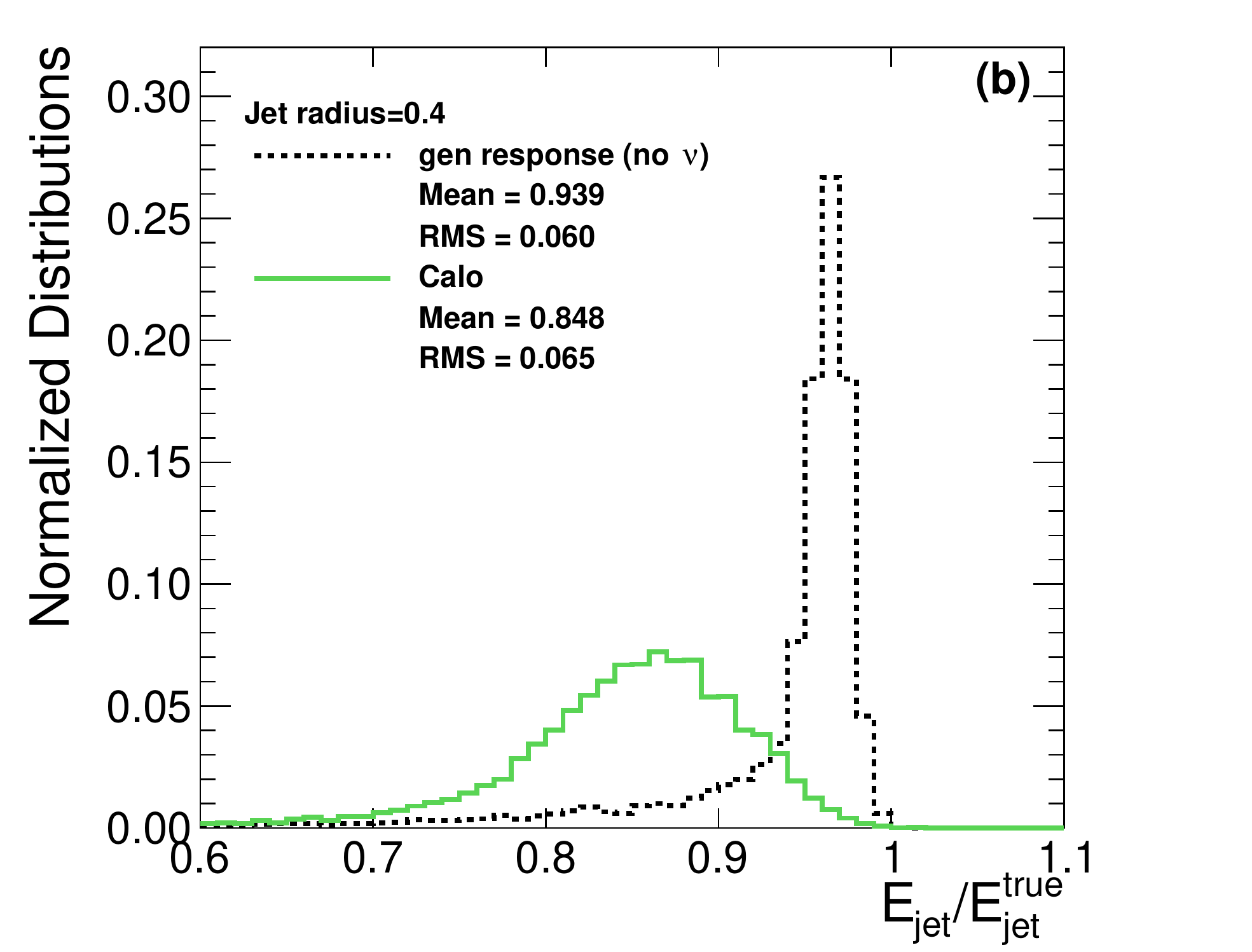} \\
\end{tabular}
\end{center}
\caption{Ratio of the weighted and smeared generator-level jet energy to 
 the true jet energy, from (a) W-jets in the 
$\mathrm{Z}^{\prime}\rightarrow \mathrm{WW}$ and from (b) Higgs-jets in 
the $\mathrm{H}_2\rightarrow \mathrm{hh}$ samples. The masses 
of the $\mathrm{Z}^{\prime}$ and $\mathrm{H}_2$ resonances 
are 10~TeV and 5~TeV, respectively. The weighted and smeared distributions 
(dashed lines) are compared with the distributions measured 
with the SiFCC calorimeters (solid lines).}
\label{fig:weight}
\end{figure}

We compare the calorimeter jet energy with the weighted 
and smeared energy of its corresponding generator-level jet. 
Within a jet, charged pions, photons, and charged kaons 
contribute $\approx 40\%$, $24\%$, and $11\%$ of the jet energy, respectively, 
while protons, neutrons, $\mathrm{K}_{s}^0$, and $\mathrm{K}_{L}^0$ contribute 
$\approx 5\%$ each. The rest of the energy comes from baryons. 
Before clustering the generator-level particles with the anti-kt algorithm, 
we modify the energy of each generator-level particle with weighing and 
smearing. %The weighing and smearing are performed as follows. 
Figure~\ref{fig:piresponse} shows a 
loss of acceptance for low-momentum tracks due to the 5-Tesla magnetic field. 
Therefore, the response of all charged particles with energy below 2~GeV is 
set to zero. The response of electrons and photons 
is set to unity and the energy is smeared using a resolution 
of $\left(0.15/\sqrt{E} \oplus 0.01\right)$. Finally, the energies
of all other charged particles are weighted and smeared 
following the response and resolution functions in 
Eqn.~\ref{eq:response}. 
Figure~\ref{fig:weight} shows the distributions of the ratio of 
weighted and smeared energy to the true jet energy from the $\mathrm{Z}^{\prime}$ 
and the $\mathrm{H}_2$ samples; the true jet energy is defined as the 
generator-level jet energy before weighing and smearing (neutrinos are 
excluded in the clustering). The weighted and smeared distributions are 
narrower than the calorimeter energy distributions; in addition, the overall 
means of the former distributions are higher. This study indicates that the 
calorimeter jet energy response and resolution can not be modeled and 
extrapolated simply from those of the single-particle samples.

\section{Energy Response and Resolution of W-jets \label{sec:jetresponse}}
\begin{figure}
\begin{center}
\begin{tabular}{cc}
\includegraphics[width=.35\textwidth]{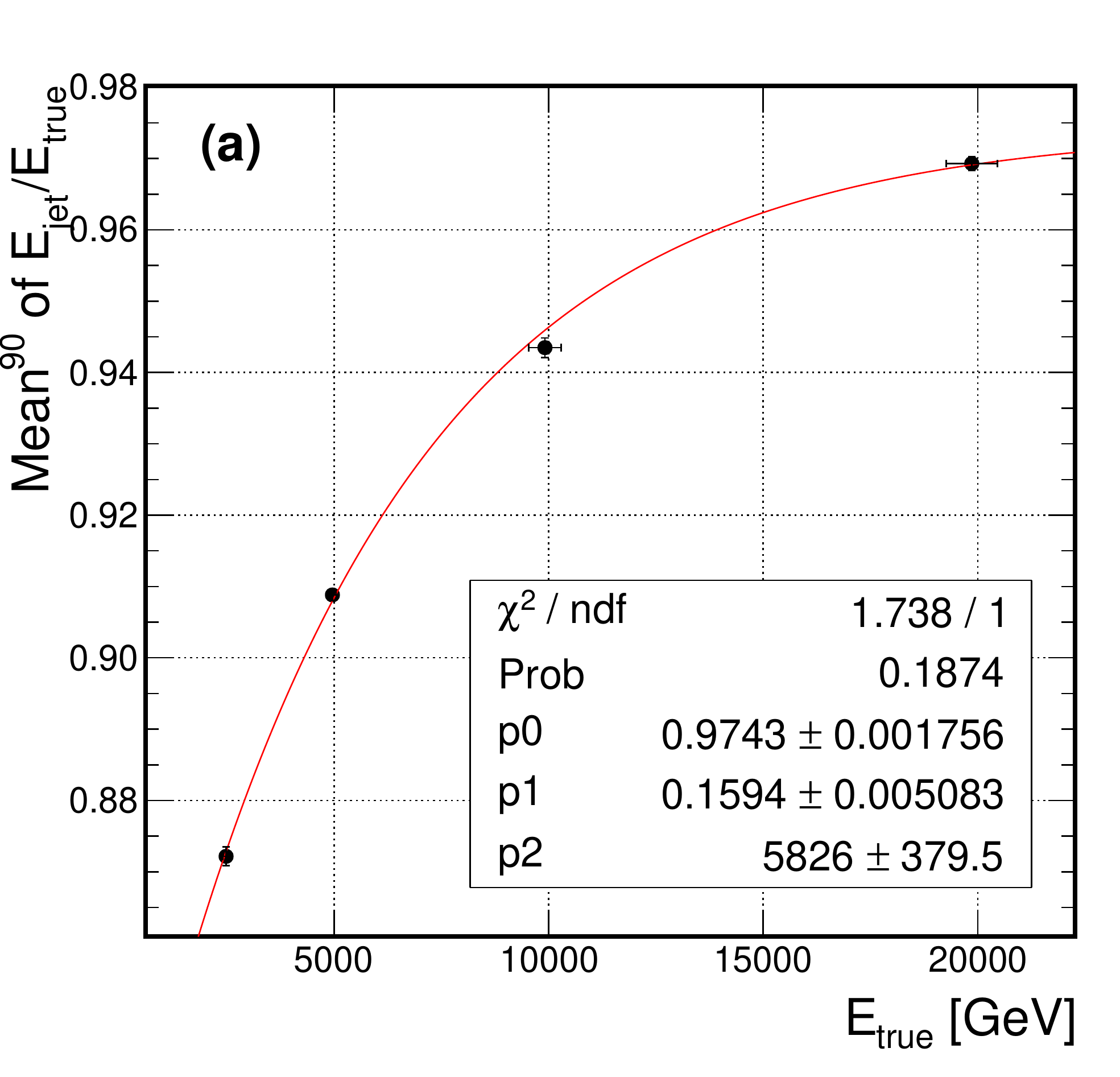} &
\includegraphics[width=.35\textwidth]{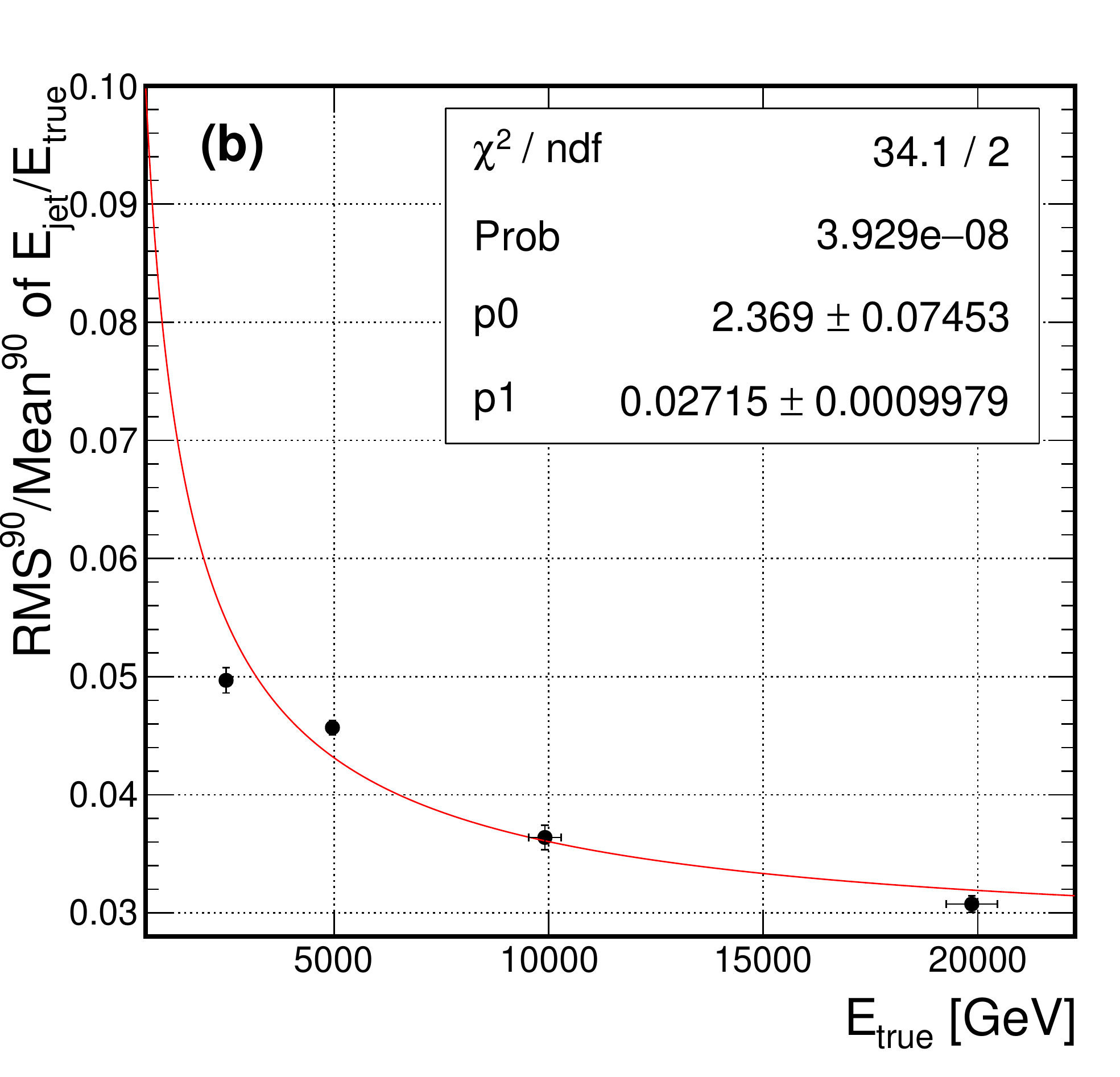} \\
\end{tabular}
\end{center}
\caption{The (a) energy response and (b) resolution of the W-jets from the 
$\mathrm{Z}^{\prime}\rightarrow \mathrm{WW}$ samples. The response is fitted 
with a function $y_\mathrm{response} = p_0\left(1-p_1 e^{-x/p_2}\right)$ 
while the resolution is fitted with a function 
$y_\mathrm{resolution}=p_0/\sqrt{E_\mathrm{true}}\oplus p_1$ as indicated 
by the curves.
}
\label{fig:wjet}
\end{figure}

We study the energy response and resolution of W-jets using the 
$\mathrm{Z}^{\prime}\rightarrow \mathrm{WW}$ samples with $\mathrm{Z}^{\prime}$ 
mass at 5--40~TeV. The $\left|\eta\right|$ of W-jets must be less than 1.1. 
Figure~\ref{fig:wjet} shows the 90\%-truncated mean and RMS/mean of the jet 
energy ratios. 
The sampling term of the W-jet energy resolution for these extreme energies 
is $\approx 237\%$ while the constant term is $\approx$2.7\%;
the sampling term is much higher than our expectation. 
%The W-jet energy 
%resolution deteriorates as the jet energy increases, which is not expected. 
However, note that this is the first time that the response and resolution of 
such high-energy jets have been studied with a full detector simulation. 
Currently, studies are ongoing to understand the reconstruction
of calorimeter clusters used for jets. 

\section{Two-Jet Separation In Boosted Bosons \label{sec:angular}}
\begin{figure}
\begin{center}
\begin{tabular}{cc}
\includegraphics[width=.35\textwidth]{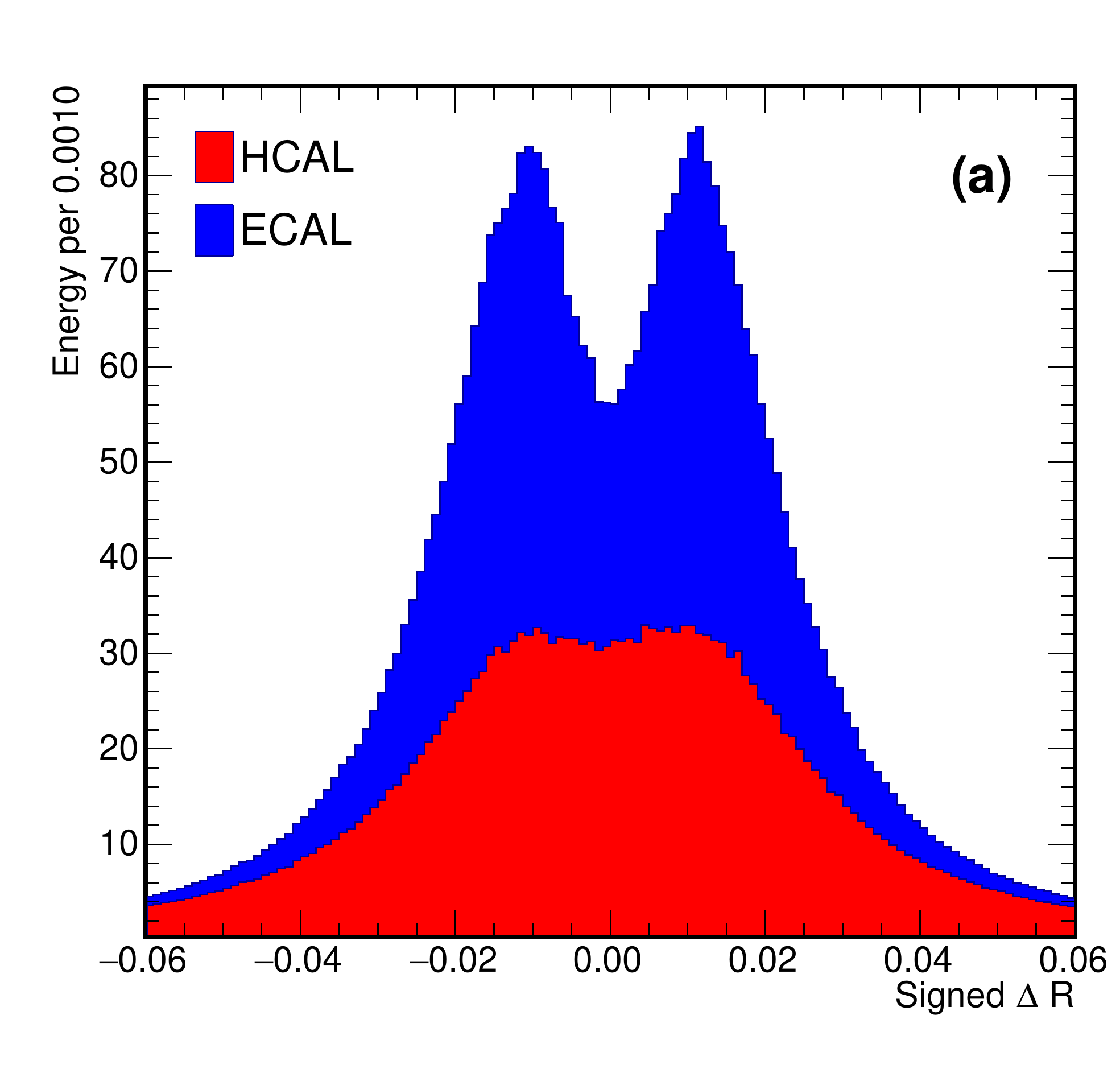} &
\includegraphics[width=.35\textwidth]{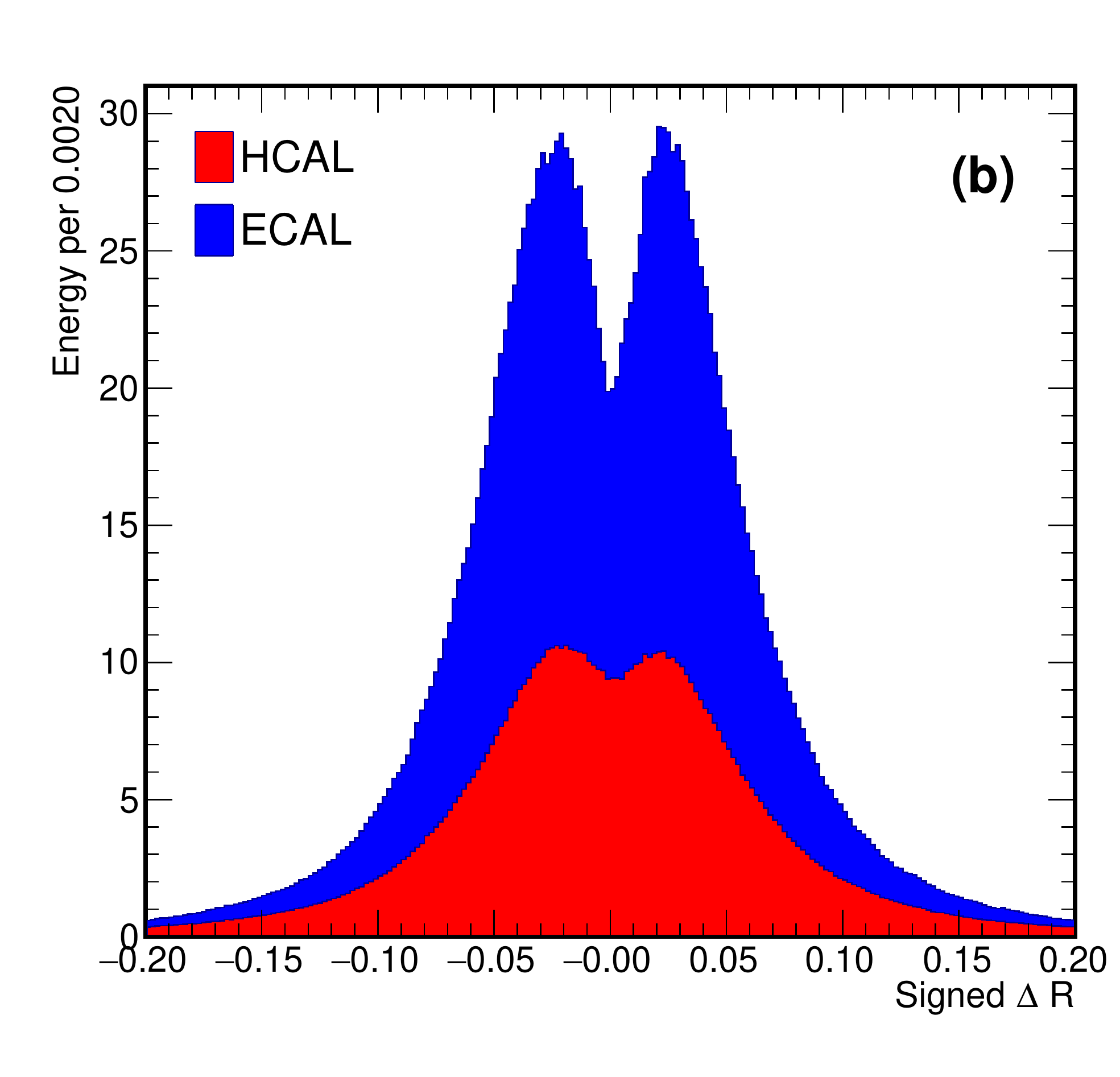}\\
\end{tabular}
\end{center}
\caption{
The ECAL and HCAL energy profiles as a function of signed $\Delta R$ from 
(a) W-jets in the $\mathrm{Z}^{\prime}\rightarrow \mathrm{WW}$ and 
from (b) Higgs-jets in the $\mathrm{H}_2\rightarrow \mathrm{hh}$ 
samples. The masses of the $\mathrm{Z}^{\prime}$ and $\mathrm{H}_2$ resonances 
are 10~TeV and 5~TeV, respectively. The separation between the two peaks in 
each distribution is $\approx 0.02$ for W-jets and $\approx 0.05$ for 
Higgs-jets. }
\label{fig:angular}
\end{figure}

In addition to the energy response and resolution, we study the angular 
separation of two jets within highly boosted bosons. The angular separation 
is quantified by a variable, signed $\Delta R$, which is defined as the 
displacement of a calorimeter hit within a W-jet (Higgs-jet), in the 
$\eta-\phi$ plane, with respect to the generator-level W (Higgs) boson-direction, projected onto the $\mathrm{q}-\bar{\mathrm{q}}$ 
axis ($\mathrm{b}-\bar{\mathrm{b}}$) : \( \mathrm{signed}~\Delta R \equiv 
\frac{\overrightarrow{D}_{\mathrm{boson,calohit}}~\bullet~\overrightarrow{D}_{\mathrm{q}\bar{\mathrm{q}}}}{\overrightarrow{D}_{\mathrm{q}\bar{\mathrm{q}}}}\). 
Figure~\ref{fig:angular} shows the ECAL and HCAL energy profiles as a function 
of signed $\Delta R$ for the W-jets and Higgs-jets. The high-granularity 
calorimeter makes it possible to see the separation of subjets within 
highly-boosted W-jets and Higgs-jets.

\section{Conclusion}
In this document, we present a first study of single particle response and 
boosted W-jets and Higgs-jets at the tens-TeV energy scale using a full 
\GEANTfour\ simulation. The 
energy response and resolution of single particles follow the expected 
performance of the designed detector while the energy resolution of 
calorimeter jets is worse than the expected values. 
The high granularity of the SiFCC calorimeters 
shows the possibility to separate energy 
clusters within a jet from a 2.5-5 TeV boosted boson. We are geared towards 
the study of high-level variables related to jet substructure and techniques 
for jet grooming.

\acknowledgments 
 This research was done using resources provided by the Open Science Grid,
which is supported by the National Science Foundation and the U.S.
Department of Energy's Office of Science. The work of SC was supported by 
the U.S. Department of Energy, Office of Science under Contract No. 
DE-AC02-06CH11357. The work of AVK was supported by the Fermi National Accelerator Laboratory.
Fermilab is operated by Fermi Research Alliance, LLC, under Contract No. DE-AC02-07CH11359 with 
the United States Department of Energy.

\end{document}